\documentclass[12pt]{article}
\usepackage{url}
\usepackage{amsmath}

\newcommand{\qt}[1]{``#1''}
\newcommand{\pp}[2]{\mbox{pp.~#1\,--\,#2}}

\begin{document}

\title{Brief History of Quantum Cryptography:\\
A Personal Perspective}

\date{17 October 2005}

\sloppy

\author{Gilles Brassard\,%
\thanks{Supported in parts by the Natural Sciences and Engi\-neering Research Council
of Canada, the Canada Research Chair Programme and the Canadian Institute
for Advanced Research.}\\
\small\sl Universit\'e de Montr\'eal\\[-0.5ex]
\small\sl D\'epartement d'informatique et de recherche op\'erationnelle\\[-0.5ex]
\small\sl C.P.~6128, Succursale Centre-Ville\\[-0.5ex]
\small\sl Montr\'eal (Qu\'ebec), H3C 3J7~~Canada\\
\small \url{http://www.iro.umontreal.ca/~brassard}}

\maketitle
\thispagestyle{empty}

\vspace{-4mm}
\begin{flushright} \small\sf
Based on my eponymous paper~\cite{self} in the\\
\textsl{Proceedings of the
IEEE Information Theory Workshop \\on Theory and Practice
in Information-Theoretic Security},\\
Awaji Island, Japan, 17 October 2005.\\
(With minor improvements on 10 April 2006)
\end{flushright}
\vspace{1mm}

\begin{abstract}

Quantum cryptography is the only approach to privacy ever proposed that
allows two parties (who do not share a long secret key ahead of time)
to communicate with provably perfect secrecy under the nose of an eavesdropper
endowed with unlimited computational power and whose technology
is limited by nothing but the fundamental laws of nature.
This essay provides a personal historical perspective on the field.
For~the sake of liveliness,
the style is purposely that of a spontaneous
after-dinner speech.
\end{abstract}

\newpage

\section{Prehistory}

The story begins in the early 1960's, when Stephen Wiesner and Charles Bennett were undergraduate students
together at Brandeis University.  Having many common friends, they enjoyed talking with
each other.  Later, after Wiesner had gone to graduate school at Columbia and Bennett at Harvard,
they kept in touch.
In~particular, the former payed frequent visits to the latter's  communal house in Boston.
During one of those visits\,\footnote{\,Probably\ldots{} Bennett's memories of those long gone days have somewhat faded.}
in the late 60's or early 70's\,\footnote{\,Ditto.}, Wiesner told Bennett of his ideas for
using quantum mechanics to
make bank\-notes that would be impossible to counterfeit according to the laws
of nature, as well as of a ``quantum multiplexing'' channel,
which would allow one party to send two messages to another
in a way that the receiving party could decide which message to read but only at the
cost of destroying the other message irreversibly.\footnote{\,It can be argued that this marks the
invention of Oblivious Transfer---or rather one-out-of-two Oblivious
Transfer~\cite{EGL}---ten years before Michael Rabin independently introduced his original concept
in the theoretical computer science community~\cite{RabinOT}.
Of~course, it took the genius of Rabin to realize the central place
that Oblivious Transfer was destined to occupy in the sky of cryptography.}
\looseness=+1

Wiesner submitted his paper ``Conjugate Coding''
to the \emph{IEEE Transactions on Information Theory}.%
\footnote{\,It is written ``Submitted to IEEE, Infor\-mation Theory'' on top of Bennett's surviving copy
of the original typewritten manuscript, but it is probably
safe to guess that the \emph{Transactions} were meant.}
Unfortunately, it was rejected, probably deemed incomprehensible by the editors and referees\,%
\footnote{\,This is an educated guess.} because it was written in the technical language
of physicists (which must have seemed normal for a physicist!).
It~is fortunate that Wiesner had expounded his ideas to Bennett,
for they might otherwise have been lost forever.
Instead, Bennett mentioned them occasionally to various people in the subsequent years,
invariably meeting with very little sym\-pathy until\ldots

\newpage

\section{History}

One fine afternoon in  late October 1979, I was swimming at the beach of a posh hotel in
San Juan, Puerto Rico. Imagine my surprise when this complete stranger swims up to me and
starts telling me, without apparent provocation on my part,
about Wiesner's quantum banknotes!
This was probably the most bizarre, and certainly the most magical,
moment in my professional life.\footnote{\,At the risk of taking some
of the magic away, I~must confess that it was not by accident that
Bennett and~I were swimming at the same beach in Puerto Rico.
We~were both there for the 20th \emph{Annual IEEE Symposium on the
Foundations of Computer Science}.  Bennett approached me because
I~was scheduled to give a talk on relativized cryptography~\cite{oracle} on the last
day of the Symposium and he thought I might be interested in Wiesner's ideas.
By~an amazing coincidence,
on my way to San Juan,  I~had read Martin Gardner's account~\cite{gardner} of
Bennett's report~\cite{CHB-omega} on Chaitin's Omega, which had just appeared in the November 1979
``Mathematical Games'' column of \emph{Scientific American}---so,
I~knew the name but I~could not recognize
Bennett in that swimmer because I did not know what he looked like.}
Within hours, we~had found ways to mesh Wiesner's coding scheme with
some of the then-new concepts of public-key cryptography~\cite{DH}.
Thus was born a wonderful collaboration that was to spin out
quantum teleportation~\cite{telep}, entanglement distillation~\cite{distil},
the first lower bound\,\footnote{\,We~were trying
to prove that unstructured search
requires linear time even on quantum computers but eventually had to settle on
our best finding, which was an $\Omega(\sqrt{n}\,)$ lower bound.
This result was rejected from all major theoretical computer science conferences
until, giving up on conferences, we published it in
\emph{SIAM~Journal on Computing}~\cite{BBBV}.
Then, Lov Grover discovered his celebrated algorithm~\cite{Grover}
to solve the same unstructured search problem in a matching time in $O(\sqrt{n}\,)$.
At~first blissfully unaware of our lower bound,  he too was disappointed in not being
able to do better!
This was a rare case in which an algorithm was proven optimal
before its discovery!} on the power of quantum computers~\cite{BBBV},
privacy ampli\-fi\-cation~\cite{BBR,BBCM} and, of course, quantum cryptography~\cite{self}.

The ideas that Bennett and I tossed around on the beach that day resulted in
the first paper ever published on quantum cryptography~\cite{BBBW}, indeed the paper
in which the term ``Quantum Cryptography'' was coined.  It~was presented at
\textsc{Crypto}~'82, an annual conference that had started
one year earlier.
By~a strange twist of history, our paper triggered the belated publication of Wiesner's
original paper in a special issue of the
ACM Newsletter \textsc{Sigact} News~\cite{conjcod}
that was otherwise devoted to a selection of papers from the earlier
\textsc{Crypto}~'81 conference.\footnote{\,The proceedings of that very
first \textsc{Crypto} had appeared solely as a
Technical Report of the University of California at Santa Barbara.}
Thus,  ``Conjugate Coding'' was finally disseminated
in 1983, admit\-tedly after Rabin's independent invention of his original version
of Oblivious \mbox{Transfer}~\cite{RabinOT}.

Wiesner's original banknotes, as well as our improvement, required quantum information to
be held captive in one place. This was a major practical drawback---if~any of this were ever
to become practical, which certainly seemed unlikely at the time---because, in those early days,
we were using photon polarization as the carrier of quantum information as if it were the only
choice. Strangely, it took us  a few years after meeting on the beach before we realized,
shortly after the \textsc{Crypto}~'82 conference, that God had meant photons to travel
rather than to stay put!\,\footnote{\,This is \emph{really} a shame because Wiesner's original
multiplexing channel already used polarized photons to \emph{transmit} information!}
This was the insight that made us think of using a quantum channel to transmit
confidential information.  But, however obvious it may seem now, we did not think
of quantum key distribution right away.

At first, we wanted the quantum signal to encode the transmitter's confidential message
in such a way that the receiver could decode it if no eavesdropper were present,
but any attempt by the eavesdropper to intercept the message would spoil it without
revealing any information. Any such futile attempt at eavesdropping would be
detected by the legitimate receiver, alerting him to the presence of the eavesdropper.
Since this early scheme was unidirectional, it required the
legitimate parties to share a secret key, much as in a one-time pad encryption.
The~originality of our scheme was that the same one-time pad could be reused safely
over and over again if no eavesdropping were detected.  Thus, the title of our paper
was ``Quantum Cryptography~II\,: How to reuse a one-time pad safely even if
\mbox{P\,=\,NP}\,''~\cite{BBB}.
We~submitted this paper to major theoretical computer science conferences,
such as STOC (The \emph{ACM~Annual Symposium on Theoretical Computer Science}),
but we failed to have it accepted.
Contrary to Wiesner's ``Conjugate Coding'', however, our ``Quantum Cryptography~II\,''
paper has forever remained unpublished (copies are available from the authors).%
\footnote{\,The idea of safely reusing a
one-time pad was resurrected two decades later by
Ivan Damg{\aa}rd, Thomas Pedersen and Louis Salvail~\cite{louis1,louis2},
who were totally unaware of our earlier work~\cite{BBB}.
It~must be said, however, that the scheme put forward by Bennett and me was
an unproven hack at best, whereas the new work is beautifully set on firm foundations.}

We all but forgot about this early idea in 1983, when we realized how much easier it
would be to use the quantum channel to transmit an arbitrarily long \emph{random} secret key.
If~eavesdropping were detected on the quantum channel, due to unavoidable disturbance,
the key would be thrown away; otherwise it could be used safely to transmit a sensitive
message by use of the classical one-time pad scheme. In~essence, this detour via the
one-time pad allowed us to turn nature's given eavesdropping-\emph{detection} channel
into an eavesdropping-\emph{prevention} channel.
Moreover, the new scheme was much more robust against lost photons since a random
subsequence of a random bit string is still a random bit string, albeit shorter.
We~wrote up a proposal for the 1983 \emph{IEEE~Symposium
on Information Theory} (ISIT), which was held in St-Jovite (near my hometown of Montr\'eal)
that year.  Not~only was it accepted (yeah!), but it was granted a long presentation.
The~corresponding one-page abstract~\cite{BB83}---such is the rule at ISIT---provides the
official birth certificate for \emph{Quantum Key Distribution}, which remains to this date
the most feasible of all proposed applications for the burgeoning field of
quantum information science.
So~feasible indeed that all you need to implement quantum key distribution is a  few
chocolate balls~\cite{chocolate}!  \verb+:-)+

Shortly thereafter, my good friend Vijay Bhargava
was in charge of a special session on coding and information theory for
yet another IEEE conference,
which took place in Bangalore, India\,\footnote{\,This conference was organized in parts
to celebrate the 75th anniversary of the Indian Institute of Technology.}, in December 1984.
He~invited me to give a talk on any subject of my choice, and naturally I chose
Quantum Cryptography considering how difficult it was to get these ideas published at the time.
The~resulting paper~\cite{BB84} gave its name to the  ``BB84~protocol'' even though
it had been described in detail as early as 19\emph{83} at the IEEE ISIT \emph{talk} but not
in the \emph{paper}
(how~much can you say in a one-page abstract?).  Retrospectively, it is amusing to note
that the only reason the BB84 protocol was finally published is that it had not
been submitted to the conference that printed it in its proceedings!
Thanks Vijay!

A~curious episode took place when Doug Wiedemann, having read Wiesner's
``Conjugate Coding'' in \textsc{Sigact} News, reinvented the \emph{exact} BB84
protocol in 1997, even called it ``quantum cryptography'', and published it in
\textsc{Sigact} News~\cite{Wie} as well; see also~\cite{Reinv}.

Throughout the 1980's, very few people took quantum cryptography seriously
and most people simply ignored~it.
Eventually, Bennett and~I decided we had to \emph{show them} by building a working prototype!
Bennett asked John Smolin to help with the hardware and I asked Fran\c{c}ois Bessette
and Louis Salvail to help with the software.
Essentially without any special budget allocated to the project, we were
able, in late October 1989, to establish history's first secret quantum transmission, over a staggering
distance of 32.5~centimetres, precisely on the tenth anniversary of our meeting
at the San Juan beach~\cite{BBBSS1,BBBSS2}\,!

It remains a mystery to me that this successful prototype made a world of difference to
physicists, who suddenly paid attention.  In~particular, it gave us the opportunity to publish
in \emph{Scientific American}~\cite{SciAm}. The~funny thing is that, while our theory
had been serious, our prototype was mostly a joke. Indeed, the largest piece in the
prototype was the power supply needed to feed in the order of one thousand volts
to Pockels cells, used to turn photon polarization. But~power supplies
make noise, \textsl{and not the same noise for the different voltages needed for
different polarizations}. So,~we could literally hear the photons as they flew, and zeroes
and ones made different noises. Thus, our prototype was unconditionally
secure \emph{against any eavesdropper who happened to be deaf}! \verb+:-)+

Perhaps inspired by our ``success'', gifted experimentalists\,%
\footnote{\,The list of experimentalists involved
goes beyond the scope of this historical essay and we prefer
to mention no one rather than to omit many.} began building ever
more sophisticated prototypes capable of realizing quantum key distribution
over tens of kilometres of optical fibre, some of which are even commercially
available. Line of sight experiments have also been carried out over similar distances.
Serious plans already exist to link earth-based users by a quantum
cryptographic process mediated by satellite.\footnote{\,In the ``simple'' version,
the satellite would be privy of the secret key it helped establish.
More sophisticated proposals would use quantum entanglement
to prevent this weakness. In~that case, a dishonest satellite can prevent the legitimate
parties from establishing a key at all, but it cannot fool them into believing
their key is shared and secret when it is not.}
The~possibilities are endless now that a worldwide quantum cryptographic network
is within reach of current technology.
Several decades ago, transatlantic telephone communications were carried out by
submarine cables.  The idea that ordinary people would soon communicate
through satellites was as
far fetched back then as an array of quantum communication satellites would appear today.
As~Theodore Roosevelt once said, ``The only limit to our realization of
tomorrow will be our doubts of today. Let us move forward with strong and active faith.''\,%
\footnote{\,As beautifully engraved in his Memorial in Washington, D.C.}

Far more important than the ``success'' of our prototype,
another factor played a crucial role in the newly-found interest of  physicists
for quantum cryptography in the early 1990's.
This was the re-reinvention of quantum key distribution by Artur Ekert,
and the fact that he published his ideas in \emph{Physical Review Letters}~\cite{E91}
rather than computer science journals and conferences.
Contrary to Wiedemann, however, Ekert did not reinvent the BB84 protocol.
Instead, he involved quantum entanglement~\cite{EPR}
and the violation of Bell's theorem~\cite{bell}.
Even though Ekert's \mbox{proposal} was equivalent to BB84 in a sense\,\cite{BBM},
the idea proved to be very fertile, particularly after the inven\-tion of
entanglement distillation~\cite{distil} and quantum privacy amplification~\cite{QPA}.
Thinking about entanglement-based cryptography has made it possible
to give much simpler proofs of unconditional security for the original un-entangled
BB84 scheme~\cite{Shor-Presk}.
Moreover, as mentioned already, the use of entanglement could be instrumental
in a truly secure satellite-based implementation of quantum cryptography.
Finally, whenever the long-term storage of quantum infor\-ma\-tion will become feasible,
entanglement-based cryptography will provide
a form of key distribution that would remain secure not only against eavesdropping,
but also against burglary.

\section{Beyond Key Distribution}

Many people think that quantum cryptography and quantum key distribution are
one and the same.
Nothing could be farther from the truth.
Let~us first recall the two results presented in Wiesner's original ``Conjugate Coding'',
which started the entire field.
Whether or not his quantum banknotes can be considered ``cryptography''
is a matter of taste, but there can be no doubt that his quantum multiplexing channel
was mainstream cryptography ahead of its time.

One of my most vivid memories took place
when our ideas for sending confidential messages by way of quantum signals
were being rejected.
I~had been visiting Bennett at his house in Croton-on-Hudson.\footnote{\,I cannot remember if this episode
took place before or after the invention of the BB84 protocol, but I think it was before.}
As~my stay was coming to an end, he drove me back to the train station.
I~boarded the train; he was facing me on the platform, waving goodbye.
And~suddenly, we realized that Wiesner's multiplexing channel could be used to
implement Rabin's oblivious transfer!
Bennett was so \mbox{excited} that he started jumping up and down as the doors closed
between~us.
As~the train picked up speed, I saw his spectacles flying off into the~air.
In~retro\-spect, I~have seen him more excited only once in the
quarter-of-century I~have known him: when we 
invented quantum teleportation one decade later.

You may wonder what the fuss was about. Didn't I~write in the first paragraph
of this essay (footnote~3) that Wiesner deserved credit for inventing oblivious transfer?
Well, at the time, only Rabin's original version of oblivious transfer had
appeared in the theoretical computer science community~\cite{RabinOT}.
It~was already regarded as a powerful primitive, although its universality for
secure two-party computation was yet to be established~\cite{kilian}.
With hindsight, it is obvious that Rabin's oblivious transfer
and Wiesner's multiplexing channels are similar,
and that the latter can serve to implement the former.%
\footnote{\,In fact, they are equivalent~\cite{equivalence}.}
Yet, the connection had not
occurred to us until that precise instant.
Out~of the blue, we realized for the first time
that quantum mechanics could help solve mainstream (classical)
cryptographic problems beyond the mere transmission of confidential information.
The sky was the limit\,\footnote{\,Well, not exactly because Wiesner had explained in
``Conjugate Coding'' how to defeat his own quantum multiplexing channel!
But the attack was very complicated and we were hoping at the time
that  it would be unfeasible to mount or possible to circumvent.},
which did not prevent Bennett's eyeglasses from shattering on the train platform.

The other cryptographic task that we studied in the early days
was \emph{coin-tossing}.
The classical concept had been pioneered by Manuel Blum
at \textsc{Crypto}~'81~\cite{coin-flip}.
It~is a little-known fact that
the 1983 ISIT abstract that introduced quantum key distribution~\cite{BB83}, as well
as its better known 1984 big brother~\cite{BB84} that gave its name to the BB84 protocol,
were just as much about quantum coin-tossing as they were about quantum key distribution.
Consider for instance the following excerpt from the original 1983 version:
\begin{quote}
We also present a protocol for coin-tossing by exchanges of quantum messages.
[\ldots] our protocol is secure against
traditional kinds of cheating, even by an opponent with unlimited computational power.
Ironically, however, it can be subverted by use of a still subtler quantum phenomenon,
the so-called Einstein-Podolsky-Rosen [EPR] paradox, which in effect allows the initiating
party to toss the coin after the responding party has announced his guess of the outcome.
\end{quote}

This was a rare instance of a paper that introduces a new protocol and breaks it in the same paper\,%
\footnote{\,Wiesner's multiplexing channel provides another such example.},
but I~guess everything is allowed in an \emph{invited} paper.  Thanks again Vijay!
I~remember when I was giving talks in the mid 1980's.
I~used to spend most of my time on quantum key distribution, because that was the ``serious''
part, but I had much more fun with coin-tossing.
This pleasure came in particular from the need to explain the EPR ``paradox''~\cite{EPR} to audiences
of computer scientists who had never heard about such a bizarre thing in those days.
Also, we were particularly proud of having discovered a way to make ``practical'' use
of entanglement, albeit for an evil purpose.
To~this day, I~still think our 1983 ISIT paper marks the first explicit use of EPR
long-distance correlations
for information processing purposes.

Nearly every time I gave this talk, someone would suggest a way to fix the coin-tossing protocol,
so as to get around the EPR attack.
Invariably, I~could find within seconds a way to adapt the attack to the modified protocol\ldots{}
until Claude Cr\'epeau came along.
Soon, Cr\'epeau joined the select group of believers, at a time when Bennett and~I were possibly
the only two active researchers in quantum cryptography.
We~quickly realized that the quantum coin-tossing protocol could easily be turned into
a quantum bit commitment protocol, which is possibly a stronger primitive.
But of course, the quantum bit commitment protocol derived from the ISIT83 and BB84 papers
were just as vulnerable to an EPR attack as the corresponding coin-tossing protocol.

Helped by Richard Jozsa and Denis Langlois, we thought at some point that we had designed
an unbreakable quantum bit commitment scheme~\cite{BCJL}.
This was a time of high hopes because we had already developed a quantum protocol to achieve
oblivious transfer provided bit commitment were available~\cite{skubi}.\footnote{\,There
are strong theoretical reasons to believe that secure oblivious transfer cannot be built
out of bit commitment in the classical world~\cite{Imp-Rud}.  This provides a definite
advantage to the quantum setting. In~particular, it makes it possible to derive
computationally secure quantum oblivious
transfer from one-way functions, a feat thought to be classically impossible.}
Alas, we had fallen into the same trap that had doomed the earlier ISIT83/BB84
coin-tossing/bit-commitment protocol~\cite{Dom,LC}\,!
After many failed attempts at fixing the problem, it was discovered that unconditionally secure
quantum bit commitment is in fact impossible~\cite{Dom2,LC2}.

Does it mean that the benefits of quantum information for cryptography cannot go further than
allowing two people to exchange messages with absolute confidentiality?
Certainly not!
Consider coin-tossing again.
Perfect coin-tossing is possible even classically if relativity is invoked~\cite{micali}.
At~a very precise moment in time\,\footnote{\,We assume that the two parties are in the
same inertial reference frame.}, each party sends a random bit to the other.
The parties have agreed that the coin-toss is heads if the bits are the same
and tails otherwise.
Cheating is not possible since relativity forbids either party to know what the other party
has sent at the time transmission is required.

The situation changes drastically if we restrict our attention to (more usual) protocols in which parties
take turn in sending messages to each other.
In~the case of classical messages, no matter the protocol, it is always possible for one of the two
parties to decide the outcome of the coin-toss given enough computational power!
Moreover, the availability of quantum computers would make such \mbox{cheating}
possible even in practice
for most proposed implementations.  In~the spirit of quantum key distribution,
can quantum mechanics help in designing protocols that remain secure against cheaters
endowed with unlimited computational power and whose technology
is limited by nothing but quantum mechanics?
The bad news is that perfect coin-tossing protocols---in which neither party could influ\-ence the
coin-toss at all---cannot exist even if the players are allowed to transmit quantum messages.
However, Andris Ambainis has discovered a wonderful quantum protocol~\cite{andris} in which
neither party can select a desired outcome and influ\-ence the process in a way
that this wish will come true with a probability better than~75\%.
This is not perfect, but it is certainly better than anything classically \mbox{achievable}.
\looseness=+1

And what about quantum bit commitment, oblivious transfer and other post-key-distribution
cryptographic tasks?
As~I~write these lines, Cr\'epeau are others are hard at work, finding ever more imaginative
ways to resurrect them  in our quantum world.

But this is where I put down my pen for tonight.

\section*{Acknowledgement}
I am most grateful to Charles Bennett, whose assistance with fact-checking
was essential, especially in the prehistory part of this essay.
In~addition, he gave me his advice on several points.
Tongue-in-cheek, he pointed out that I should mention that I~did \emph{not} accept
all his suggestions for otherwise the resulting \mbox{paper} would have been more sedate but boring!
This paper owes its existence to the fact that Ueli Maurer invited me
to participate in the \textit{IEEE Information Theory Workshop on Theory
and Practice in Information-Theoretic Security},
\mbox{Awaji} \mbox{Island}, Japan.
Furthermore, I~acknowledge the help of my students
Anne Broadbent and Andr\'e Allan M\'ethot in the final stretch of the
\mbox{writing} and submission process
to the Proceedings of that Workshop~\cite{self}.

\end{document}